# Stabilization of vector soliton complexes in nonlocal nonlinear media

Zhiyong Xu, Yaroslav V. Kartashov,* and Lluis Torner

*ICFO-Institut de Ciencies Fotoniques, and Universitat Politecnica de Catalunya,*

*Mediterranean Technology Park, 08860 Castelldefels (Barcelona), Spain*

We introduce vector soliton complexes in nonlocal Kerr-type nonlinear media. We discover that under proper conditions the combination of nonlocality and vectorial coupling has a remarkable stabilizing action on multi-humped solitons. In particular, we find that stable bound states featuring several field oscillations in each soliton component do exist. This affords stabilization of vector soliton trains incorporating large number of humps, a class of structures known to self-destroy via strong instabilities in scalar settings.

The mutual interaction between nonlinear excitations is under active consideration in different areas of physics, including condensed matter, dynamical bio-molecules, solid-state physics, and nonlinear optics. This includes a rich variety of effects connected with the vectorial nature of the nonlinear excitations. For example, vectorial coupling was observed in multicomponent Bose-Einstein condensates [1], and in topological defects arising due to the interspecies interaction [2]. In nonlinear optics, vectorial coupling between several light waves resulting in formation of vector solitons has been extensively studied for coherent [3] and incoherent [4,5] interactions in materials with local nonlinearity. One important result introduced in Ref [4] is the existence of multi-humped solitons afforded by the vectorial interactions. Recently, vector solitons in media with transverse periodic modulation of refractive index were addressed as well [6]. It was shown [3-6] that vectorial coupling in local Kerr-type media could lead to existence of complex multi-humped structures that have no counterpart in the scalar case. However, in a local saturable medium such structures were found to be made stable only when the total intensity distribution *does not feature more than three* humps [5].

On the other hand, nonlinearity may be highly nonlocal, a property that drastically alters the propagation and interaction of nonlinear waves [7]. Nonlocality should be taken into account when the transverse extent of the wave-packet becomes comparable with the characteristic response length of the medium. The nonlocal nonlinear response brings new features in development of modulation instability [8], it prevents catastrophic collapse of multidimensional solitons [9,10], and results in stabilization of vortex solitons [11]. New effects attributed to nonlocality have been studied in photorefractive crystals [12], thermo-optical materials [13], liquid crystals [14], plasmas [15], and Bose-Einstein condensates with long-range inter-particle interactions [16]. Among other specific features encountered with nonlocal medium is the possibility of formation of complex multi-humped trains composed of scalar bright or dark solitons [17]. Such trains may or may not feature threshold number of humps for stability depending on the type of nonlocal response of the medium. In particular, in materials with exponential response functions stable scalar trains cannot involve more than four solitons [18].

In this Letter we introduce vector soliton structures in nonlocal media, and reveal that such combination affords remarkable new phenomena. In particular, we discovered that because of the nature of soliton interactions in nonlocal media, vector solitons can form stable bound states that feature *several field oscillations* in *each* component, in sharp contrast to the behavior encountered in local media. We reveal that nonlocal nonlinear response plays a strong stabilizing action for vector solitons of higher orders. Complex patterns with large number of humps in one field component, that are unstable when propagating alone, can be made stable in nonlocal media due to the mutual coupling with stable solitons propagating in other field component.

We consider the propagation of two mutually incoherent laser beams along the $\xi$ axis of a nonlocal focusing Kerr-type medium described by the system of equations for dimensionless complex light field amplitudes $q_{1,2}$ and nonlinear correction to refractive index $n$ given by:

$$i\frac{\partial q_1}{\partial \xi} = -\frac{1}{2}\frac{\partial^2 q_1}{\partial \eta^2} - q_1 n,$$
$$i\frac{\partial q_2}{\partial \xi} = -\frac{1}{2}\frac{\partial^2 q_2}{\partial \eta^2} - q_2 n, \qquad (1)$$
$$n - d\frac{\partial^2 n}{\partial \eta^2} = |q_1|^2 + |q_2|^2.$$

Here $\eta$ and $\xi$ stand for the transverse and longitudinal coordinates scaled to the input beam width and diffraction length, respectively; the parameter $d$ describes the degree of nonlocality of the nonlinear response. In the limit $d \to 0$ the system (1) reduces to a system of coupled nonlinear Schrödinger equations for the fields $q_{1,2}$ whose vector soliton solutions are well established [3-5]. The opposite limit $d \to \infty$ corresponds to strongly nonlocal regime. Under appropriate conditions the mathematical model (1) adequately describes nonlinear response of some thermo-optical materials, liquid crystals or partially ionized plasmas [13-15]. Among the conserved quantities of system (1) are the energy flows $U, U_{1,2}$ and the Hamiltonian $H$:

$$U = U_1 + U_2 = \int_{-\infty}^{\infty} (|q_1|^2 + |q_2|^2) d\eta,$$

$$H = \int_{-\infty}^{\infty} \left[ \frac{1}{2} \left| \frac{\partial q_1}{\partial \eta} \right|^2 + \frac{1}{2} \left| \frac{\partial q_2}{\partial \eta} \right|^2 - \frac{1}{2} (|q_1|^2 + |q_2|^2) \int_{-\infty}^{\infty} G(\eta - \lambda)(|q_1|^2 + |q_2|^2) d\lambda \right] d\eta, \quad (2)$$

where $G(\eta) = (1/2d^{1/2}) \exp(-|\eta|/d^{1/2})$ is the response function of the nonlocal medium.

We searched for the stationary solutions of Eq. (1) numerically in the form $q_{1,2}(\eta,\xi) = w_{1,2}(\eta) \exp(ib_{1,2}\xi)$ where $w_{1,2}(\eta)$ are real functions and $b_{1,2}$ are real propagation constants. The resulting system of equations obtained after substitution of light field in such form into Eq. (1) was solved with a standard relaxation method. To elucidate the linear stability of the solutions, we searched for perturbed solutions in the form $q_{1,2}(\eta,\xi) = [w_{1,2}(\eta) + u_{1,2}(\eta,\xi) + iv_{1,2}(\eta,\xi)] \exp(ib_{1,2}\xi)$, where real $u_{1,2}(\eta,\xi)$ and imaginary $v_{1,2}(\eta,\xi)$ parts of the perturbation can grow with a complex rate $\delta$ upon propagation. Linearization of Eq. (1) around a stationary solution $w_{1,2}$ yields the eigenvalue problem

$$\delta u_1 = -\frac{1}{2} \frac{d^2 v_1}{d\eta^2} + b_1 v_1 - n v_1,$$

$$\delta v_1 = \frac{1}{2} \frac{d^2 u_1}{d\eta^2} - b_1 u_1 + n u_1 + w_1 \Delta n,$$

$$\delta u_2 = -\frac{1}{2} \frac{d^2 v_2}{d\eta^2} + b_2 v_2 - n v_2, \quad (3)$$

$$\delta v_2 = \frac{1}{2} \frac{d^2 u_2}{d\eta^2} - b_2 u_2 + n u_2 + w_2 \Delta n,$$

where $\Delta n = 2\int_{-\infty}^{\infty} G(\eta - \lambda)[w_1(\lambda)u_1(\lambda) + w_2(\lambda)u_2(\lambda)]d\lambda$ is the perturbation of refractive index. We solved Eq. (3) numerically to find the profiles of perturbations and the associated growth rates.

The simplest vector solitons can be found with $b_1 = b_2$ in the form $w_1(\eta) = w(\eta)\cos\varphi$ and $w_2(\eta) = w(\eta)\sin\varphi$, where $w(\eta)$ describes the profile of scalar soliton, and $\varphi$ is an arbitrary projection angle. The most interesting situation is encountered, however, when $b_2 \neq b_1$ and the first and second soliton components possess different types of symmetry. Below, without loss of generality, we search for solutions with $b_2 \leq b_1$. The properties of vector soliton composed from the first nodeless and second dipole-mode components are summarized in Fig. 1. At fixed propagation constant $b_1$ and nonlocality degree $d$ there exist lower $b_2^{\text{low}}$ and upper $b_2^{\text{upp}}$ cutoffs on $b_2$ for vector soliton existence. As $b_2 \to b_2^{\text{low}}$ the second dipole-mode component gradually vanishes (Fig. 1(a)), while in the opposite limit $b_2 \to b_2^{\text{upp}}$ the nodeless component ceases to exist (Fig. 1(b)). Such transformation of the internal structure of vector soliton is accompanied by the development of two-humped refractive index distribution near $b_2^{\text{upp}}$. Notice, that with increase of nonlocality degree, the width of refractive index distribution increases substantially and far exceeds the width of the actual intensity distribution $w_1^2 + w_2^2$. The deep on top of the refractive index distribution becomes more pronounced in local limit ($d \to 0$) and almost vanishes in strongly nonlocal medium ($d \to \infty$). At small degrees of nonlocality and at $b_2 \to b_2^{\text{upp}}$ vector soliton transforms into two very well separated solitons with bell-shaped second components having opposite signs, while at $b_2 \to b_2^{\text{low}}$ small second component broadens substantially in comparison with the localized first component. In strongly nonlocal media both components remain well localized in the cutoffs. The total energy flow $U$ is found to be a monotonically increasing function of $b_2$ (Fig. 1(c)). The energy sharing $S_{1,2} = U_{1,2}/U$ as a function of $b_2$ is depicted in Fig. 1(d). We found that the width of the existence domain on $b_2$ for vector solitons shrinks substantially with increase of nonlocality degree $d$ (Fig. 1(e)) and expands with increase of $b_1$ (Fig. 1(f)). A comprehensive linear stability analysis revealed that vector solitons composed of nodeless and dipole-mode components are stable in the entire domain of their existence.

We also found vector solitons composed of first nodeless and second triple-mode components, whose properties are summarized in Fig. 2. The generic properties of such solitons are reminiscent to those of solitons discussed in Fig. 1, but the existence domain for such solitons is substantially wider (compare Figs. 2(c) and 1(e)). The linear stability analysis revealed that solitons composed from nodeless and triple-mode components are unstable for small values of $d$. With decrease of $d$ the instability domain emerges near upper cutoff $b_2^{\text{upp}}$ and occupies the entire domain of soliton existence as $d \to 0.2$ (Fig. 2(c)). In contrast, at moderate nonlocality degrees $d \sim 2.3$ such solitons are stable near lower and upper cutoffs and feature only narrow instability band inside their existence domain. At small values of $d$ the instability is of exponential type while for $d > 2.3$ we encountered only oscillatory instabilities. It should be pointed out that, in contrast to the case of local saturable medium [5], multi-humped vector solitons in nonlocal media may be stable when total intensity and refractive index distributions develop three or even more humps.

One of the central results of this Letter is that vector solitons were found to form stable bound states that feature *several field oscillations* in *both* components and that, to the best of our knowledge, were not encountered previously in any model of local Kerr-type media. Such bound states exist because of specific character of soliton interactions in nonlocal medium, whose sign is determined not only by the phase difference, but also by the separation between solitons [17].

Thus, in nonlocal media both $w_1$ and $w_2$ can change their sign in contrast to the case of local media where one of the components remains nodeless [3-5]. Several representative examples are shown in Fig. 3, including solitons incorporating dipole-mode first and triple- (Fig. 3(a)) or quadrupole-mode (Fig. 3(b)) components, as well as more complex structures (Figs. 3(c),(e),(f)). The total number of humps in the refractive index distribution is determined by the relative weights of the components. While borders of the existence domain for bound states (e.g., Figs. 3(a),(b)) look qualitatively similar to those shown in Figs. 1 and 2, the complexity of the structure of stability/instability domains increases progressively with increase of the number of humps in each component, so that typically several stability windows appear in the $(d, b_2)$ plane.

To confirm the outcome of the linear stability analysis, we performed numerical simulations of Eq. (1) with the input conditions $q_{1,2}(\eta, \xi=0) = w_{1,2}(\eta)[1 + \rho_{1,2}(\eta)]$, where $w_{1,2}(\eta)$ stands for the profiles of stationary solitons, and $\rho_{1,2}(\eta)$ are random functions with Gaussian distribution and variance $\sigma_{\text{noise}}^2$. Numerical simulations confirmed the results of the linear stability analysis in all cases. Stable solitons shown in Figs. 1-3 propagate over huge distances, exceeding experimentally feasible nonlinear material lengths by several orders of magnitude, even in the presence of considerable broadband input noise (Fig. 4).

Another important result that we encountered is that complex patterns with large number of humps, that are unstable when propagating alone, can be made stable in suitable parameter regions due to the mutual coupling with stable single- or multi-humped components. Illustrative examples are shown in Figs. 3(c),(e) and 3(f), where stabilization of five- and six-hump components (that are unstable when propagating alone in a medium with exponential response function $G$ [18]) is achieved because of the coupling with first stable fundamental and stable quadrupole-mode components, respectively. To stress that *stabilization of such complex multi-* *humped structures* takes place in a *wide parameter regions* we show in Fig. 3(d) the stability/instability domains for solitons depicted in Figs 3(c) and 3(e). Besides expected stability domain which adjoins the lower cutoff $b_2^{\text{low}}$ where vector soliton bifurcates from stable fundamental scalar soliton, another stability domain that appears at $d \approx 3$ (shaded area in the inset of Fig. 3(d)) was found. In this latter domain, stabilization of five-hump component is achieved at expense of coupling with weak fundamental component. For soliton profile belonging to this stability domain, see Fig. 3(e). Dropping the first component causes fast splitting and decay of the second component (Fig. 4(f)) that is stable in the presence of coupling (Fig. 4(e)). Similar results were obtained for more complex solitons, e.g., Fig. 3(f). Therefore, vectorial coupling of *unstable* solitons in one field component with *stable* solitons in other field component allows substantial increase of the number of solitons that can be packed into the composite, *stable* vector soliton complex.

We thus conclude stressing that vectorial coupling in Kerr-type nonlocal media, introduced here for the first time, features important new soliton phenomena. In particular, we revealed that in such media vector solitons could form stable bound states that exhibit several field oscillations in each component, thus affording an extended number of peaks that can be packed into stable soliton trains. Our predictions open a route to the experimental observation of such multi-humped soliton complexes. We based our analysis on a general model that combines nonlocal nonlinearity and vectorial coupling, potentially relevant to a variety of multicomponent nonlinear excitations in strongly nonlocal materials with symmetric nonlocal kernels. Formation of stable vector soliton complexes with rich internal structure might be a feature possible in several of such systems. Therefore, our results motivate specific research in settings that exhibit diverse mechanisms resulting in nonlocal nonlinearities, including charge transport in photorefractive crystals [12], thermal nonlinearities [13], reorientational nonlinearities in liquid crystals [14], many-body interactions in multi-species Bose-Einstein condensates [16,19] and partially ionized plasmas [15], nonlinearities of semiconductors [20], or interatomic interactions in strongly correlated atomic gases [21].

*On leave from Physics Department, M. V. Lomonosov Moscow State University, Russia. This work has been partially supported by the Generalitat de Catalunya, by the Government of Spain through grant TEC2005-07815, and by the Ramon-y-Cajal program.

# Figure captions

Figure 1.  Soliton profiles corresponding to $b_1 = 3$, $d = 2$, and (a) $b_2 = 1.8$, (b) $b_2 = 2.5$. (c) Energy flow versus propagation constant $b_2$ at $b_1 = 3$. Points marked by circles correspond to solitons shown in (a) and (b). (d) Energy sharing between $w_1$ and $w_2$ soliton components versus $b_2$ at $b_1 = 3$, $d = 2$. Domains of existence of vector solitons at $(d,b_2)$ plane for $b_1 = 3$ (e) and at $(b_1,b_2)$ plane for $d = 2$ (f).

Figure 2.  Soliton profiles corresponding to $b_1 = 3$, $d = 2$, and (a) $b_2 = 1$, (b) $b_2 = 2.3$. Domains of existence of vector solitons at $(d,b_2)$ plane for $b_1 = 3$ (c) and at $(b_1,b_2)$ plane for $d = 2$ (d).

Figure 3.  Profiles of various higher-order vector solitons at $b_2 = 2.55$, $d = 2$ (a), $b_2 = 1.5$, $d = 2$ (b), $b_2 = 0.45$, $d = 4$ (c), $b_2 = 2.15$, $d = 10$ (e), and $b_2 = 2.37$, $d = 16$ (f). Panel (d) shows stability (shaded) and instability domains on $(d,b_2)$ plane for solitons incorporating nodeless $w_1$ and five-hump $w_2$ components (see panels (c) and (e) for soliton profiles). In all cases $b_1 = 3$.

Figure 4.  Propagation dynamics of vector solitons in nonlocal medium. (a) Soliton involving nodeless $w_1$ and dipole-mode $w_2$ components at $b_1 = 3$, $b_2 = 2$, $d = 2$. (b) Soliton involving nodeless $w_1$ and triple-mode $w_2$ components at $b_1 = 3$, $b_2 = 2.4$, $d = 3$. (c) Soliton involving dipole-mode $w_1$ and triple-mode $w_2$

components at $b_1 = 3$, $b_2 = 2.55$, $d = 2$. (d) Soliton involving dipole-mode $w_1$ and quadrupole-mode $w_2$ components at $b_1 = 3$, $b_2 = 1.5$, $d = 2$. (e) Soliton involving nodeless $w_1$ and five-hump $w_2$ components at $b_1 = 3$, $b_2 = 0.45$, $d = 4$. (f) Decay of soliton depicted in (e) in the absence of $w_1$ component. In (a)-(e) white noise with variance $\sigma_{noise}^2 = 0.01$ is added into input stationary profiles. Only $w_2$ components are shown.

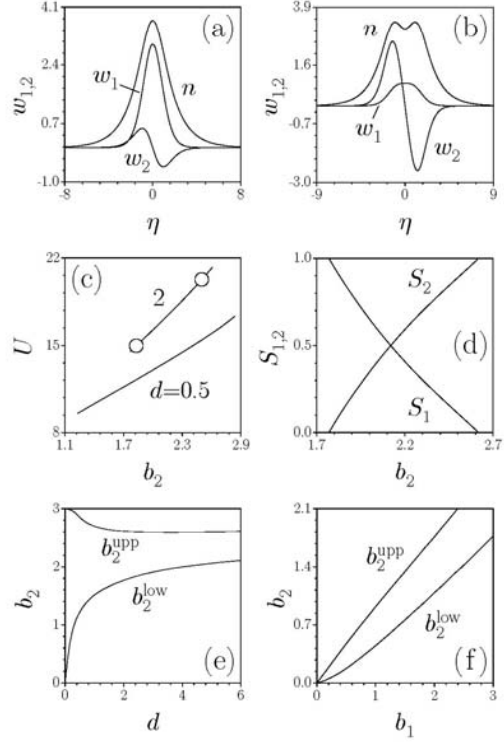

Figure 1. Soliton profiles corresponding to $b_1 = 3$, $d = 2$, and (a) $b_2 = 1.8$, (b) $b_2 = 2.5$. (c) Energy flow versus propagation constant $b_2$ at $b_1 = 3$. Points marked by circles correspond to solitons shown in (a) and (b). (d) Energy sharing between $w_1$ and $w_2$ soliton components versus $b_2$ at $b_1 = 3$, $d = 2$. Domains of existence of vector solitons at $(d, b_2)$ plane for $b_1 = 3$ (e) and at $(b_1, b_2)$ plane for $d = 2$ (f).

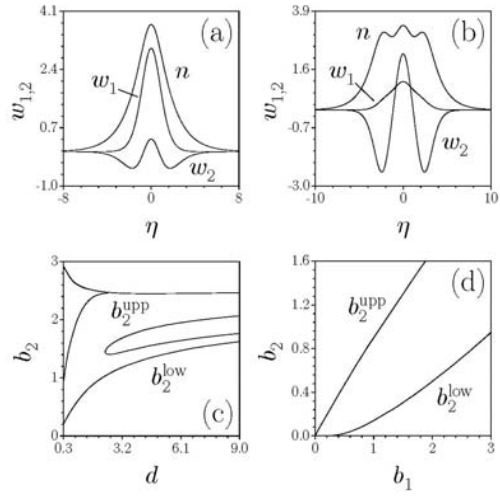

Figure 2. Soliton profiles corresponding to $b_1 = 3$, $d = 2$, and (a) $b_2 = 1$, (b) $b_2 = 2.3$. Domains of existence of vector solitons at $(d, b_2)$ plane for $b_1 = 3$ (c) and at $(b_1, b_2)$ plane for $d = 2$ (d).

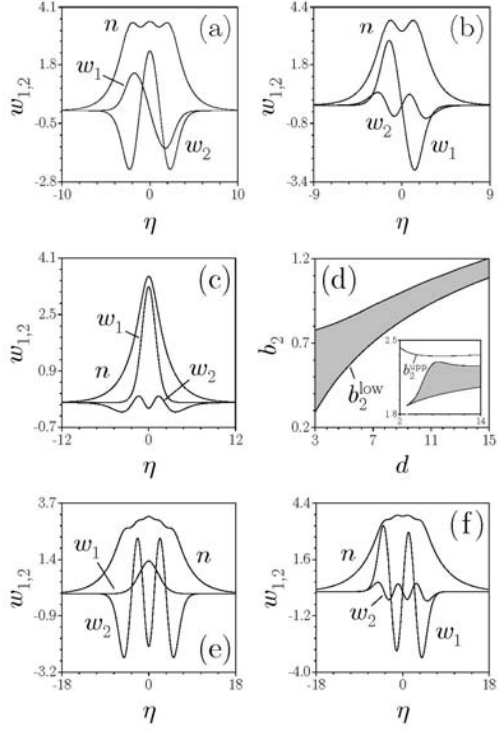

Figure 3. Profiles of various higher-order vector solitons at $b_2 = 2.55$, $d = 2$ (a), $b_2 = 1.5$, $d = 2$ (b), $b_2 = 0.45$, $d = 4$ (c), $b_2 = 2.15$, $d = 10$ (e), and $b_2 = 2.37$, $d = 16$ (f). Panel (d) shows stability (shaded) and instability domains on $(d, b_2)$ plane for solitons incorporating nodeless $w_1$ and five-hump $w_2$ components (see panels (c) and (e) for soliton profiles). In all cases $b_1 = 3$.

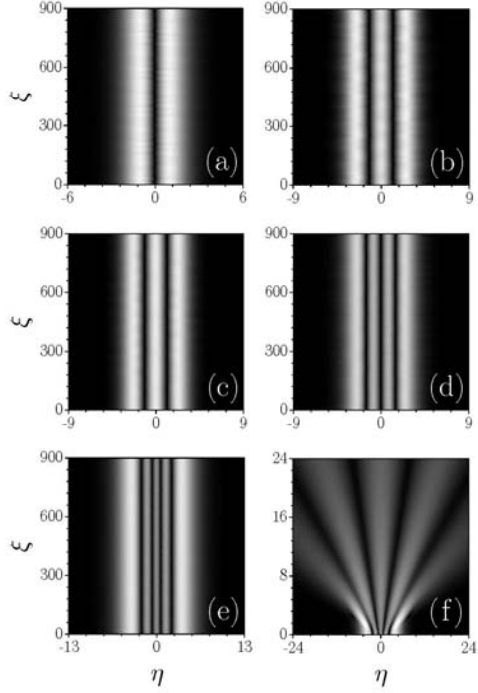

Figure 4. Propagation dynamics of vector solitons in nonlocal medium. (a) Soliton involving nodeless $w_1$ and dipole-mode $w_2$ components at $b_1 = 3$, $b_2 = 2$, $d = 2$. (b) Soliton involving nodeless $w_1$ and triple-mode $w_2$ components at $b_1 = 3$, $b_2 = 2.4$, $d = 3$. (c) Soliton involving dipole-mode $w_1$ and triple-mode $w_2$ components at $b_1 = 3$, $b_2 = 2.55$, $d = 2$. (d) Soliton involving dipole-mode $w_1$ and quadrupole-mode $w_2$ components at $b_1 = 3$, $b_2 = 1.5$, $d = 2$. (e) Soliton involving nodeless $w_1$ and five-hump $w_2$ components at $b_1 = 3$, $b_2 = 0.45$, $d = 4$. (f) Decay of soliton depicted in (e) in the absence of $w_1$ component. In (a)-(e) white noise with variance $\sigma_{\text{noise}}^2 = 0.01$ is added into input stationary profiles. Only $w_2$ components are shown.